**FUNGAR: a pipeline for detecting antifungal resistance mutations directly from metagenomic short reads**

Short title: FUNGAR: antifungal resistance in metagenomes

Henrique RM Antoniolli[1,2], Lívia Kmetzsch[1,2,3] & Charley C Staats[1,2,3,*]

[1]ResGen – One Health Antimicrobial Resistance Genomic Surveillance Initiative.

[2]Departamento de Biologia Molecular e Biotecnologia, Instituto de Biociências, Universidade Federal do Rio Grande do Sul, UFRGS, Brazil.

[3]Programa de Pós-graduação em Biologia Celular e Molecular, Centro de Biotecnologia, Universidade Federal do Rio Grande do Sul, UFRGS, Brazil.

*Corresponding author. Mailing address: Avenida Bento Gonçalves 9500, Prédio 43421, Centro de Biotecnologia, Bloco 4, Campus do Vale, UFRGS. Caixa Postal 15005. Porto Alegre, RS, Brasil, 91501-970. Phone 55 51 3308 6080; Fax 55 51 3308 7309. Email: staats@ufrgs.br. ORCID: 0000-0002-2433-6903

**Abstract**

**Motivation:** Antifungal resistance has become an increasing global concern in both clinical and environmental health. Detecting known resistance mutations directly from sequencing reads, in special metagenomic samples, remains a major challenge. As fungal pathogens are often neglected compared with bacterial pathogens, most available tools are designed for bacterial taxa, whereas tools targeting fungi typically require assembled genomes. In metagenomic datasets, assembly-based strategies may result in substantial information loss due to genome fragmentation, low-abundance species, or incomplete recovery of resistance loci.

**Results:** Here, we present FUNGAR, an open-source pipeline for the rapid identification of antifungal resistance genes and mutations directly from short-read data. FUNGAR employs translated alignments with DIAMOND and curated data from the FungAMR database to detect amino acid substitutions across all six open reading frames. The pipeline produces structured, reproducible reports linking detected variants to their associated antifungal drugs and can be easily customized for new species or databases.

**Availability and implementation:** FUNGAR is available at https://github.com/resgen-br/fungar

## 1 Introduction

Global emergence and spread of resistance to antifungal drugs represents an increasing threat to both human and environmental health (van Rhijn and Rhodes 2025). Cases of resistance to azoles, echinocandins, and polyenes are frequently reported, especially in clinically relevant genera such as *Aspergillus*, *Candida* and *Cryptococcus* (Lockhart *et al.* 2023). Molecular mechanisms underlying such resistance phenotypes are, in most cases, well characterized and often produced by mutations in specific genes, hereafter named antifungal resistance genes (AfRGs). Those are mostly related to the alteration in the binding sites of drugs, alteration in cell wall components and nucleic acid synthesis or repair, as well as the production of biofilm (Czajka *et al.* 2023; Lee *et al.* 2023).

Despite the wide availability and the increasing ease of obtaining high throughput sequencing data, identifying AfRGs in metagenomic samples is still challenging. As demonstrated for bacteria (Maguire *et al.* 2020; Abramova *et al.* 2024), assembly-based strategies may lead to loss of information due to fragmentation of low-abundance genomes, chimeric contigs, or incomplete recovery of target loci (Mirete *et al.* 2025). Most of the existing software or tools were designed for bacterial taxa, and either focus on assembled single genomes (Bédard *et al.* 2025) or are not easily customizable for fungi species (Matsumura *et al.* 2025).

Here, we present FUNGAR, an open-source pipeline for the rapid identification of AfRGs and associated mutations in metagenomic samples. FUNGAR is based on translated alignments with DIAMOND and curated resistance data from the FungAMR database (Bédard *et al.* 2025) to detect amino acid substitutions across the six open reading frames. The pipeline is easily customizable and provides reports linking detected variants to their associated antifungal drugs.

## 2 Software description

FUNGAR is a pipeline written in Bash and Python 3 languages and is designed for Unix-like operating systems (built and tested on Ubuntu 24.04.3 LTS). The only required input files are either quality-filtered paired- or single-end FASTQ files. Reads are then aligned against protein sequences and their respective mutations described in the database FungAMR (Bédard *et al.* 2025) with the blastx mode in DIAMOND (Buchfink *et al.* 2021), which screens for alignments in all six possible open reading frames (ORFs). By default, the minimum length of a translated sequence (-orf) is set to 50 amino acids (for read length of 150bp) for increased specificity. The user can also set custom values for minimum query coverage (--min-query-cover), minimum percent of identity (--min-pident), and the genetic code used for query translation (-code). The overall pipeline flow is outlined in Figure 1.

The output files are then processed with the *pandas* library by comparing in both reference and query sequences the amino acid present in the known position of mutations described in the database FungAMR (Bédard *et al.* 2025); detected variants are cross-referenced with their associated antifungal drugs, and results are compiled into a structured CSV file. This output includes the sequencing read ID, mutation details (gene, position, reference/variant amino acids), and the corresponding drug resistance profile. Results are also summarized into a CSV file containing the gene, position, reference/variant amino acids, drug resistance profile, and number of reads supporting each mutation.

The pipeline includes prebuilt DIAMOND databases for fungal species commonly found in environmental and clinical samples, which will be continuously updated. Users may also supply custom databases by specifying a directory (via the -d flag) containing two files: (1) a DIAMOND-formatted protein database (*.dmnd); and (2) a CSV file listing resistance mutation with columns for gene, position, reference, mutation, and fungicide. This flexibility allows the pipeline to adapt to novel species or emerging resistance mechanisms.

## 3 Validation with synthetic dataset

The functionality of FUNGAR was first evaluated with a manually built dataset of FASTQ files. Two mutations were introduced in the second residue of the dihydrofolate reductase (*DHFR*) protein sequence of *Pneumocystis jirovecii* (NCBI accession number ABB84736.1): (i) D2E, which confers resistance to methotrexate (Bédard *et al.* 2024); and (ii) D2F, a neutral mutation with no documented resistance phenotype. The amino acid sequences were then reverse translated with EMBOSS 6.6.0 *backtranseq* (Rice *et al.* 2000). Eight paired-end reads (150 bp) were constructed covering the first 300 nucleotides—the first four reads with the mutations in the forward strand, while the remaining four with the mutations in the reverse complement strand (Table S1). All base calls were assigned a Phred quality score of 30 to simulate high-quality sequencing data. The pipeline successfully detected only the D2E mutation described in the database. Furthermore, it was able to detect mutations in different ORFs, and in both forward and reverse complement nucleotide sequences (Table S1).

**4 Application to real datasets**

In the second round of tests, FUNGAR was applied to both a genomic dataset (Rhodes *et al.* 2022) and a metagenomic dataset (de Almeida *et al.* 2020), hereafter referred to as datasets G and M, respectively. Dataset G consisted of ten environmental isolates of *Aspergillus fumigatus*, eight of which were confirmed as azole-resistant (itraconazole, posaconazole and voriconazole) by MIC analysis (accession numbers in Table S2). Dataset M included 11 samples from three cystic fibrosis patients (named as A, B and C) colonized by fungi, primarily *Aspergillus* and *Candida* species (accession numbers in Table S3). Raw FASTQ files from all datasets were quality filtered using fastp (Chen *et al.* 2018). For dataset M, reads that concordantly aligned to the *Homo sapiens* reference genome (GCF_000001405.40) were removed using Bowtie2 (Langmead and Salzberg, 2012). The final number of reads after these filtering steps is reported in Tables S2 and S3. FUNGAR was then run using the *A. fumigatus* database for dataset G, while for dataset M, separate databases were created for *Aspergillus* and *Candida* to include all relevant genes and mutations.

FUNGAR successfully detected in dataset G the mutation L98H in the *Cyp51A* protein, which provides resistance to several azole drugs, as described by Rhodes *et al.* (2022). Several other AfRGs that were not included in their datasets but are described in FungAMR (Bédard *et al.* 2025) were detected by FUNGAR—*Fks*, *Hmg1*, *Hmg2*, and *Sdh,* which confer resistance to caspofungin, micafungin, and amphotericin B (Table S2). Our pipeline also detected mutations in other AfRGs (*Fks*, *Hmg1* and *Hmg2*) in samples that were considered by those authors as wildtype (samples C95 and C96). As for dataset M, AfRGs were detected in the three patients, across a total of five samples. In patient A, the same mutation (F200Y) was identified in the beta-tubulin gene of *A. fumigatus* in two out of five samples, conferring resistance to azoles. In patient B, a single AfRG was detected, corresponding to the H270R mutation in the *Sdh* gene of *A. fumigatus*, which is also associated with azole resistance. Finally, patient C harbored the same *Sdh* H270R mutation, as well as the D153E mutation in the *Erg11* gene of *Candida*, both of which confer resistance to azoles.

## 5 Conclusions

To our knowledge, FUNGAR is the first pipeline that detects and annotates known mutations in AfRGs directly from short-read sequencing data. This pipeline aids the detection of AfRGs in metagenomic samples—and may be employed in genomic samples as well—and provides a fast, reproducible, and extensible framework that may assist the monitoring of emerging antifungal resistance mechanisms.

**Conflict of interest**

None declared.

**Funding**

This work was supported by Conselho Nacional de Desenvolvimento Científico e Tecnológico (CNPq) [grant no. 408717/2022-0, and postdoctoral fellowship no. 383432/2024-3 to HRMA], Ministério da Ciência, Tecnologia e Inovação (MCTI) and Fundo Nacional de Desenvolvimento Científico e Tecnológico (FNDCT), as part of the *ResGen*–One Health Antimicrobial Resistance Genomic Surveillance Initiative.

## Data availability

The data underlying this article are available on Zenodo and can be accessed under the following doi: 10.5281/zenodo.17612490.

**Figure legends**

**Figure 1.** General workflow of FUNGAR. Single- or paired-end reads are aligned against a protein database, containing amino acid sequences of known genes that confer resistance to antifungal drugs (AfRGs), using DIAMOND on blastx mode in all six possible open reading frames. Alignments are cross-referenced against known mutations in AfRGs, and resistance profiles are then summarized on a CSV file.

**Supplementary data**

**Table S1** – Description of mutations inserted in manually built FASTQ files for the dihydrofolate reductase (*DHFR*) protein sequence of *Pneumocystis jirovecii* and the result obtained with FUNGAR.

**Table S2** – Results obtained with FUNGAR after screening genomic data (Rhodes *et al.* 2022) for *Aspergillus fumigatus*.

**Table S3** – Results obtained with FUNGAR after screening metagenomic data for *Aspergillus* and *Candida*.

Table S1

| Read pair | Mutation | Gene strand | ORF | Detection |
|---|---|---|---|---|
| 1 | D2E | Forward | 1 | yes |
| 2 | D2F | Forward | 1 | no |
| 3 | D2E | Forward | 2 | yes |
| 4 | D2F | Forward | 2 | no |
| 5 | D2E | Reverse | 1 | yes |
| 6 | D2F | Reverse | 1 | no |
| 7 | D2E | Reverse | 2 | yes |
| 8 | D2F | Reverse | 2 | no |

Table S2

| Sample | SRA | N reads | Protein | Mutation | Fungicide |
|---|---|---|---|---|---|
| C85 | ERR9791654 | 6.823.764 | Beta-tubulin | F200Y | Azole(s) |
| | | | Cyp51A | L98H | Azole(s) |
| | | | Fks | S53G | Caspofungin;Micafungin |
| | | | Hmg1 | E105K | Isavuconazole;Itraconazole; Posaconazole;Voriconazole |
| C86 | ERR9791655 | 7.000.574 | Beta-tubulin | E198A | Azole(s) |
| | | | Cyp51A | L98H | Azole(s) |
| | | | Fks | S53G | Caspofungin;Micafungin |
| | | | Hmg2 | I235S | Amphotericin_B;Isavuconazole; Itraconazole;Posaconazole;Voriconazole |
| | | | Sdh | H270Y | Azole(s) |
| C87 | ERR9791656 | 7.635.498 | Fks | S53G | Caspofungin;Micafungin |
| C88 | ERR9791657 | 7.183.820 | Beta-tubulin | F200Y | Azole(s) |
| | | | Cyp51A | L98H | Azole(s) |
| | | | Fks | S53G | Caspofungin;Micafungin |
| | | | Hmg1 | E105K | Isavuconazole;Itraconazole; Posaconazole;Voriconazole |
| C89 | ERR9791658 | 6.961.510 | Beta-tubulin | F200Y | Azole(s) |

| Sample | SRA | N reads | Protein | Mutation | Fungicide |
|---|---|---|---|---|---|
| | | | Fks | S53G | Caspofungin;Micafungin |
| | | | Hmg1 | E105K | Isavuconazole;Itraconazole;Posaconazole;Voriconazole |
| C91 | ERR9791659 | 6.513.176 | Beta-tubulin | F200Y | Azole(s) |
| | | | Fks | S53G | Caspofungin;Micafungin |
| C92 | ERR9791660 | 7.062.116 | Cyp51A | L98H | Azole(s) |
| | | | Fks | S53G | Caspofungin;Micafungin |
| | | | Hmg1 | E105K | Isavuconazole;Itraconazole;Posaconazole;Voriconazole |
| | | | Sdh | H270Y | Azole(s) |
| C93 | ERR9791661 | 6.387.266 | Beta-tubulin | F200Y | Azole(s) |
| | | | Hmg1 | L273F | Itraconazole;Voriconazole |
| | | | Sdh | H270Y | Azole(s) |
| C95 | ERR9791662 | 6.926.710 | Fks | S53G | Caspofungin;Micafungin |
| | | | Hmg1 | W272C | Isavuconazole;Itraconazole;Posaconazole;Voriconazole |
| | | | Hmg1 | W272L | Isavuconazole;Itraconazole;Posaconazole;Voriconazole |

| Sample | SRA | N reads | Protein | Mutation | Fungicide |
|---|---|---|---|---|---|
| | | | Hmg2 | I235S | Amphotericin_B;Isavuconazole; Itraconazole;Posaconazole;Voriconazole |
| C96 | ERR9791663 | 7.269.154 | Fks | S53G | Caspofungin;Micafungin |
| | | | Hmg1 | E306K | Amphotericin_B,Caspofungin,Itraconazole,Posaconazole,Voriconazole |

Table S3

| Sample | SRA | N reads | Species | Gene | BLAST | Mutation | Fungicide |
|---|---|---|---|---|---|---|---|
| PA1 | SRR12149839 | 379.752 | *Aspergillus fumigatus* | Beta-tubulin | Beta-tubulin (KY212756) | F200Y | Azole(s) |
| PA2 | SRR12149836 | 278.366 | Not found | | | | |
| PA3 | SRR12149835 | 270.446 | Not found | | | | |
| PA4 | SRR12149830 | 1.046.486 | Not found | | | | |
| PA5 | SRR12149837 | 3.465.740 | *Aspergillus fumigatus* | Beta-tubulin | Beta-tubulin (XM_013473589) | F200Y | Azole(s) |
| PB1 | SRR12149834 | 291.300 | Not found | | | | |
| PB2 | SRR12149829 | 1.727.610 | *Aspergillus fumigatus* | *Sdh* | *Sdh* (XM_007306866) | H270R | Azole(s) |
| PC1 | SRR12149840 | 428.890 | Not found | | | | |
| PC2 | SRR12149833 | 540.716 | Candida sp. | *Erg11* | *Erg11* (NM_001179137) | D153E | Fluconazole;Itraconazole;Voriconazole |
| PC3 | SRR12149832 | 422.290 | Not found | | | | |
| PC4 | SRR12149831 | 1.545.904 | *Aspergillus fumigatus* | *Sdh* | *Sdh* (PV330449.1) | H270R | Azole(s) |